\documentclass[usenatbib]{mn2e} \usepackage{graphicx,subfigure,times}
\usepackage{times,graphicx}

\def\R200{\ensuremath{R_{\mathrm{200}\ }}}

\newcommand{\nm}{\mbox{\ensuremath{\mathrm{~\nm}\ }}}

\newcommand{\km}{\mbox{\ensuremath{\mathrm{~km}\ }}}

\newcommand{\kpc}{\mbox{\ensuremath{\mathrm{~kpc}}}}
\newcommand{\Mpc}{\mbox{\ensuremath{\mathrm{~Mpc}}}}
\newcommand{\s}{\mbox{\ensuremath{\mathrm{~s}\ }}}

\newcommand{\GHz}{\mbox{\ensuremath{\mathrm{~GHz}\ }}}
\newcommand{\mJy}{\mbox{\ensuremath{\mathrm{~mJy}\ }}}

\newcommand{\pMpc}{\ensuremath{\mathrm{\Mpc^{-1}\ }}}
\newcommand{\ps}{\ensuremath{\mathrm{\s^{-1}\ }}}

\newcommand{\kmps}{\ensuremath{\mathrm{\km \ps}\ }}

\newcommand{\mJpbeam}{\ensuremath{\mathrm{mJy/beam\ }}}

\title[Jet speeds in WATs]{Jet speeds in wide angle tailed radio
  galaxies} \author[N. N. Jetha et al]{Nazirah
  N. Jetha${^1}$\thanks{Email: nnj@star.sr.bham.ac.uk}, Martin J. Hardcastle${^2}$, and Irini Sakelliou${^3}$ \\
  ${^1}${School of Physics and Astronomy, University of Birmingham,
    Edgbaston, Birmingham B15 2TT}\\ ${^2}${School of Physics,
    Astronomy and Mathematics, University of Hertfordshire, College
    Lane, Hatfield, Hertfordshire AL10 9AB}\\${^3}${Max Planck
    Institut f\"{u}r Astronomie, K\"{o}nigstuhl 17, D-69117,
    Heidelberg}}

\begin{document}

\label{firstpage}

\maketitle

\begin{abstract}
We present a sample of 30 wide angle tailed radio galaxies (WATs) that
we use to constrain the jet speeds in these sources.  We measure the
distribution of jet-sidedness ratios for the sample, and assuming that
the jets are beamed, jet speeds in the range (0.3-0.7)c are obtained.
Whilst the core prominence of the sample, which ought to be a reliable
indicator of beaming, shows little correlation with the jet-sidedness,
we argue that due to the peculiar nature of WATs core-prominence is
unlikely to be a good indicator of beaming in these sources.  We
further show that if the jets are fast and light, then the galaxy
speeds required to bend jets into C-shapes such as those seen in
0647+693 are reasonable for a galaxy in a merging or recently merged cluster.  
\end{abstract}

\section{Introduction}
\label{introduction}

Usually classified as FRI sources \citep{fr1974}, wide angle tailed
radio galaxies (WATs) are now thought to be an interesting hybrid of
both FRI and FRII morphologies; they exhibit well collimated inner
jets, which flare suddenly into diffuse plumes.  These jets, which
typically extend for tens of kiloparsecs and often terminate in
compact radio-bright features similar to the hotspots of FRII sources
\citep{HS2003}, appear superficially similar to the longer jets seen
in FRII sources, whereas the diffuse plumes resemble more those found
in FRI sources.  The causes of the sudden jet flaring have been
explored in detail in \citet{Jethaa}, where we note that the
transition from a cooler core of X-ray emitting gas to a hotter, extended
intra-cluster medium (ICM) co-incides with the jet flaring.  However,
as noted in our conclusions, there is uncertainty as to the nature of
the jets; FRII jets are in general tightly collimated, and have
moderate to high jet speeds \citep[e.g.][]{Hardcastle1999}, whereas
FRI jets tend to have large opening angles and also tend to be slower
than FRII jets \citep[e.g.][]{1999MNRAS.306..513L}.  Whilst jets in
WATs superficially look like those of FRII sources, in that WAT jets
tend to be well collimated, and have polarisation structures that
resemble those found in FRII sources \citep{HS2003}, it is
unclear whether their speeds resemble those of FRII sources or FRI
sources.  Establishing the nature of the WAT jets would allow us to
establish whether WATs are failed FRIIs
\citep{1999AA...349..381H}, or if they are a different sort of
radio source altogether.  Further, understanding the nature of the
jets may help to explain the dynamics of the jets in clusters, in
particular why the jets flare suddenly, and the relationship between
jet length and the cluster environment.

The question of WAT jet speeds has been investigated before by
\citet{ODonoghue}, who use a Monte Carlo analysis of a sample
of 10 WAT sources to obtain a jet speed of $0.2c$.  \citet{HS2003}
also address the issue of speeds in WAT jets, and obtain jet speeds of
\(0.3c\).  In both cases, the sources chosen for the samples
investigated tend to exhibit two sided jets and this may have biased
the results in the direction of reduced beaming effects and lower jet speeds.

In this paper, we use a larger sample of WATs, drawn from the
catalogue of Abell clusters of \citet{OLVII}, to better constrain the
jet speed in WATs. We describe our sample selection procedure and data in
Section~\ref{sample}; in Section~\ref{speeds}, we discuss how we infer
jet speeds from our data, and the speeds inferred; and in
Section~\ref{discussion}, we discuss the implications of this
result.  Throughout the paper, we assume that \(H_0\)=70\kmps\( \!
\!\)\pMpc\(\!\) and use J2000 co-ordinates.

\section{Sample selection and observations}
\label{sample}

An initial sample of 67 sources was selected from the maps of the
\citet{OLVII} 21-cm survey of Abell clusters.  Our selection criteria
were that a source had to be included in the \citet{OLVII} sample, and
that there had to be evidence of the possibility of there being two
well collimated jets, which was inferred from the existence of two
bright plume bases aligned with the core, and lying approximately
equidistant from the core.  Sources with very one-sided jets were not
excluded from the sample, and no account was taken of the distance of
the source from the cluster centre.  Sources where the inner jets are
very bent, like 0647+693 \citep[see][]{HS2003} were not excluded from
the sample.  Well known WAT objects, such as Hydra A, which were
excluded from the \citet{OLVII} sample, were also excluded from this
sample to keep our sample homogeneous.

The VLA data archive was then searched for high frequency, high
resolution data for the sample objects.  In cases where no archival
data of the quality required were available, 8-\GHz VLA observations
were requested, and the sources were observed as part of this study.
Those sources are marked in Table~\ref{sampledetails}, and details of
these observations and unpublished radio maps, both from the new
observations and archival data, are presented in Appendix~\ref{maps}.
None of the radio sources observed by ourselves was excluded from the
WAT sample; they all exhibited WAT morphology; however, two of the sources
were excluded from this analysis, as the jets were not detected. 

\begin{figure}
\scalebox{.4}{\includegraphics{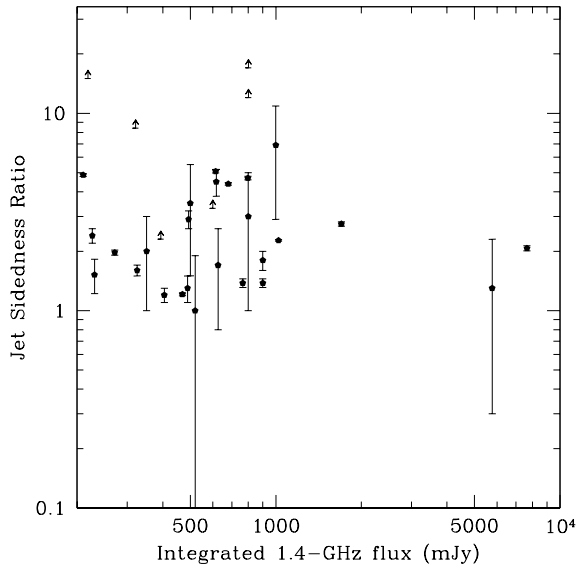}}
\caption{The jet sidedness ratios of the sources in our sample plotted
  against $S_{1400}$ \protect{\citep[taken from][]{OLVII}}.  The
  filled points represent actual jet detections, whilst the {\bf {arrows}} show lower limits on the sidedness ratio.}
\label{samplefig}
\end{figure}

The data for all 67 candidate sources (both archival and our own
observations) were reduced in {\sc aips} in the standard way.  After
examination of the more detailed images provided by this process,
sources that did not exhibit WAT morphology were excluded from the
sample.  We excluded 33 sources from the sample for this reason, and a
further 4 (including the two from our own observations) were excluded
from this analysis due to there not being enough information in the
current maps to define the straight parts of the jets, resulting in a sample of 30 sources,
which is presented in Table~\ref{sampledetails} and
Fig~\ref{samplefig}.  For each source the core flux was measured using
the {\sc aips} task {\sc jmfit}.  The jet and counter-jet fluxes,
which form the basis of the jet speed estimate, were also measured,
using a combination of {\sc aips} scripts that rotate the radio image
so that the jets are horizontal, and allow the straight part of the
jet and corresponding counter-jet regions to be defined, before the
flux in those regions is measured, and corrected for the background.
An error for the jet/counter-jet ratio was calculated by estimating
the off source noise (in $\mathrm{mJy/beam}$) for the jet and
counter-jet, and multiplying by the square root of the number of beams
in the jet and counter-jet regions.  The errors on the jet and
counter-jet fluxes were then combined together in the usual way to
obtain an error on the jet/counter-jet ratio.

\begin{table*}
\caption{Details of the sources in our sample}
\begin{tabular}{lcccccccl}
\hline
Abell cluster & radio name & $z$      & \(S_{1400}\)& obs. frequency& core flux & jet flux\(^1\) & counter-jet flux\(^1\) & jet/counter-jet flux ratio\\
              & (B1950)    &        &   /(\mJy)     &  (\GHz)        &    (\mJy)        & (\mJy)    &(\mJy)               \\ \hline
A98           &0043+201    & 0.11   &    900      &  4.8          &12.3\(\pm\)0.04 &2.30      &1.67   &1.38$\pm$0.07\\  
A160          &0110+152    & 0.045  &    1000     & 4.8      &8.19 \(\pm\)0.05&2.02   &0.293          &6.9$\pm$4\\
A194          &3C40        & 0.018  &     5770   &  4.8      &89.7\(\pm\)0.08& 4.59  & 3.67          &1.3$\pm$1\\    
A562          & 0647+693   & 0.11   &     800     &  8.5      &4.81\(\pm\)0.03& 0.518 & 0.172         &3.0$\pm$2\\  
A595          &0745+521    &0.068   &    520      &  4.8     &8.2\(\pm\)0.2   & 2.60  & 2.55          &1.0$\pm$0.9\\
A610$^{\dag}$ &0756+272    & 0.096  &    493      & 8.5      &4.62\(\pm\)0.02 &6.37   &2.18           &2.9$\pm$0.3\\
A690         &0836+290    & 0.079  &    1022     & 4.8      &137 \(\pm\)0.05 &21.4   &0.311          &2.27$\pm$0.02\\
A761         &0908-102    & 0.092  &    177      & 4.8      &8.90\(\pm\)0.07 &5.61   &0.203          &28$\pm$2\\
A763          &0909+162    & 0.085  &    183      & 4.8      &1.13\(\pm\)0.06 &1.54   &0.196          &7.84$\pm$2\\
A950$^{\dag}$ &1011+500    & 0.21   &    394      & 8.5      &41.0 \(\pm\)0.1 &0.202  &$<$0.0880         &$>$2.3$\pm$0.3\\    
A1238         &1120+013    & 0.072  &    230      &  1.4     &4.4\(\pm\)0.4   &2.00   &1.31           &1.52$\pm$0.3\\              
A1346\(^*\)   &1138+060    & 0.097  &    469      &  1.4     &-           &3.49   &2.89           &1.21$\pm$0.02\\
A1446         &1159+583    & 0.10   &    765      &  4.8     &7.74\(\pm\)0.04 &3.51   &2.54           &1.38$\pm$0.07\\    
A1462$^{*\dag}$&1202+153    & 0.14   &    614      & 8.5      &-           &39.5   &7.77           &5.08$\pm$0.1\\
A1529$^{\dag}$&1221+615    & 0.23   &    321      & 8.5      &15.9\(\pm\)0.1  &4.94   &$<$0.587          &$>$8.42$\pm$0.08\\
A1559         &1231+674    & 0.11   &    900      &  8.5     &5.41\(\pm\)0.01 &2.65   &1.49           &1.8$\pm$0.2\\
A1684         &  1306+107A &  0.086 &     405     &   4.8     &6.42\(\pm\)0.03& 9.61  & 0.172         &1.2$\pm$0.1\\       
A1763         &1333+412    & 0.23   &    797      & 8.5      &3.23\(\pm\)0.01 &5.07   &1.09           &4.7$\pm$0.1\\
A1940         &1433+553    & 0.14   &    500      & 8.5      &14.97\(\pm\)0.01&1.70   &0.620          & 3.5$\pm$2\\
A1942         &1435+038    & 0.22   &    801      & 4.8      &4.3 \(\pm\)0.1  &4.77   &$<$0.280          &$>$17$\pm$3\\
A2029         &1508+059    & 0.077  &    489      &  8.5     &2.78\(\pm\)0.01 &1.10   &0.878          &1.3$\pm$0.2\\  
A2214         &1636+379    & 0.16   &     617     &   4.8     &6.78\(\pm\)0.06& 6.06  & 1.36          &4.5$\pm$0.7\\              
A2220         &1638+538    & 0.11   &    626      &  1.4     &10.1\(\pm\)0.5  &5.08   &2.97           &1.7$\pm$0.9\\
A2249$^{\dag}$&1707+344    & 0.081  &    680      & 8.5      &6.37\(\pm\)0.07 &16.3   &3.72           &4.39$\pm$0.06\\
A2304         &1820+689    & 0.088  &    801      & 1.4      &55.8\(\pm\)0.3  &7.00   &$<$0.599          &$>$12$\pm$3\\
A2306         &1826+747    & 0.13   &    600      & 4.8      &1.52\(\pm\)0.04 &0.262  &$<$0.0792         &$>$3.3$\pm$0.3 \\
A2372         &2142-202    & 0.058  &    351      &  4.8     &0.9\(\pm\)0.1   &0.270  &0.170          &2$\pm$1\\
A2395$^{\dag}$&2152+085    & 0.15   &    226      & 8.5      &5.79 \(\pm\)0.03&3.91   &1.62           &2.4$\pm$0.2\\
A2462         &2236-176    & 0.073  &    1700     & 8.5     &11.11 \(\pm\)0.02&17.9   &6.52           &2.75$\pm$0.07\\
A2634         &3C465       & 0.030  &    7650     & 8.5     &213.49\(\pm\)0.03&28.3   &13.8           &2.07$\pm$0.06\\
\hline

\end{tabular}
\vspace{0.2cm}
\begin{minipage}{16cm}
\small NOTES:\(^*\)We were unable to detect the core.  \(^1\)The jet
and counter-jet fluxes were measured at 8.5~\GHz.  When measurements
at 8.5~\GHz were not available, measurements were made at other
frequencies, and values for the jet fluxes were corrected assuming
\(\alpha\)=0.5 and \(S\left(\nu\right)\propto \nu^{-\alpha}\).
Redshifts and 1.4~\GHz fluxes were obtained from \citet{OLV}, and
\citet{OLVII} respectively, except for the flux of 3C40, which was
measured from our new radio map.  Sources marked with a $^{\dag}$ were new observations for the purposes of this study.
\end{minipage}
\label{sampledetails}
\end{table*}

\section{Estimating the jet speed and modelling intrinsic sidedness}
\label{speeds}

The jet part of the radio source can be modelled as two anti-parallel
streams of plasma being emitted from the core of the radio source.
The jets are inclined at some random angle \(\theta\) to the line of
sight; one jet travels towards the observer, the other travels away
from the observer.  Since the bulk velocity of the jet is relativistic,
the flux of the jet moving towards us will be brightened via Doppler
boosting, whereas the jet moving away from from the observer will have
its flux suppressed by a similar mechanism.

For a jet travelling at a speed \(v=\beta c\), the relationship
between the rest frame flux \(S_0\left(\nu\right)\) and the Doppler
boosted observed flux, \(S\left(\nu\right)\) is given by
\begin{equation}
S\left(\nu\right)=S_0\left(\nu\right)\left[\gamma\left(1-\beta\cos\theta\right)\right]^{-\left(2+\alpha\right)}
\label{doppboost}
\end{equation} where \(\gamma\) is the Lorentz factor, \(\gamma=(1-\beta^2)^{-1/2}\).

Since Eqn.~\ref{doppboost} can be applied equally to the jet and
counter-jet, the jet/counter-jet ratio \(R\), for a single source can
place a constraint on \(\beta\cos\theta\).  \(R\) is given by \citep{1985ApJ...295..358L}:
\begin{equation}
R=k\left(\frac{1+\beta\cos\theta}{1-\beta\cos\theta}\right)^{-(2+\alpha)}
\label{ratio}
\end{equation}  Here, \(k\) is the ratio between the rest frame jet and
counter-jet flux; it is generally thought to be approximately unity,
but can be varied to represent intrinsic variations in jet
sidedness. The spectral index \(\alpha\) is a measure of how the
flux of the jet varies with frequency, and is defined such that
\(S(\nu)\propto\nu^{-\alpha}\).  For the jets that are being dealt
with here, we take \(\alpha=0.5\).

Assuming that the sources are randomly distributed over $\theta$, a model
distribution of \(R\) can be generated for given values of \(\beta\)
and \(k\), which can be compared with the observed distribution to
obtain the most likely value for \(\beta\) in WATs.

We use a Monte Carlo technique to generate a distribution of jet
sidedness ratios for given values of $\beta$ and parameters affecting
the intrinsic sidedness. This is then compared with the observed
distribution to determine the values of the input parameters that
maximize the likelihood.  Rather than sampling a grid over all of
parameter space, which can be computationally intensive, we use a
Markov-Chain Monte Carlo (MCMC) algorithm to choose the model
parameters to investigate. The implementation of the MCMC algorithm is
the {\sc metro} sampler discussed by \citet{Hobson2004}, kindly
provided by Mike Hobson. After an initial burn-in period to allow the
Markov chain to converge, sets of model parameters are drawn from the
chain, so that the highest-likelihood regions of parameter space are
also the best sampled. The use of this algorithm to determine jet
speeds will be discussed in more detail by \citet{Mullin}.

It is usually assumed that there is no intrinsic difference between
the two jets, i.e. $k$ is unity.  However, the physical conditions in
both jet and counter-jet are unlikely to be identical, and this will
lead to some intrinsic sidedness between the jet and counter-jet.
Initially, we consider the case where the jet and counter-jet are
symmetrical, and constrain the beaming speed $\beta$ on the assumption
that all of the observed sidedness distribution can be explained by
beaming.  We then relax this assumption, and simulate the case where
some intrinsic asymmetry exists between the jet and counter-jet. 

Our simulated sidedness data are constructed by generating different
samples of jet/counter-jet pairs, with each sample having a different
$\beta$, and equal rest frame fluxes for each jet/counter-jet pair.
The angle that the jets make with the line of sight, $\theta$ is
generated randomly, in line with our assumption that there is
no preferential angle.  As can be seen from
  Table~\ref{sampledetails}, some of our sample have only lower limits
on the jet sidedness.  This is taken into account in the fitting code
by assigning a likelihood to the limit, such that the data point
could take any value of sidedness in the parameter space that is
greater than the limit.  When intrinsic sidedness is introduced, each
jet and counter-jet in a given sample has a different rest frame flux,
but the same $\beta$.  The rest frame flux is drawn from a truncated
Gaussian distribution, centred on unity, and the width of the
distribution ($\sigma$) is determined from the data.  Thus, $k$ is
determined from the ratio of two truncated Gaussian distributions.
The beamed sidedness distribution that is generated is then compared
with the actual data, using a maximum likelihood method to determine
which simulated sample best matches the observed distribution.

\subsection{Modelling jets with beaming only}
\label{beamingonly}

We initially modelled the case that the sidedness distribution was
caused by beaming alone.  The width of the Gaussian describing the
intrinsic sidedness was set to zero, such that $k$ in
Eqn.~\ref{doppboost} was unity.  This gave a value of jet speed that
created a sample which best matched the observed distribution.  This
jet speed was then used to generate synthetic samples firstly in order
to act as a check that the code was fitting the observed distribution
correctly, and secondly to obtain a measure of the dispersion in the
obtained value of the jet speed, which gives us an estimate of the
uncertainty on the fitted parameters.

\begin{figure}
\subfigure[]{{\label{3a}}\scalebox{.4}{\includegraphics[angle=270]{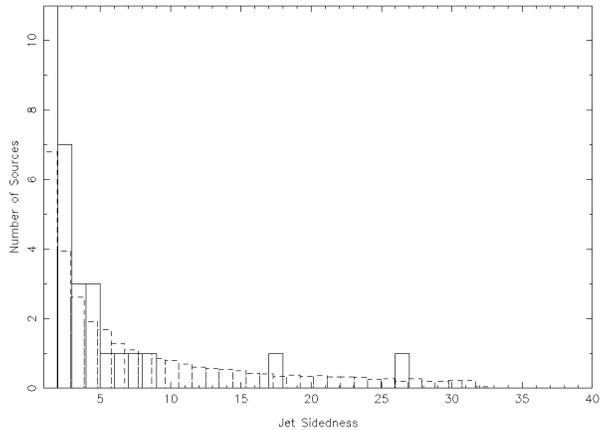}}}
\subfigure[]{{\label{3b}}\scalebox{.4}{\includegraphics[angle=270]{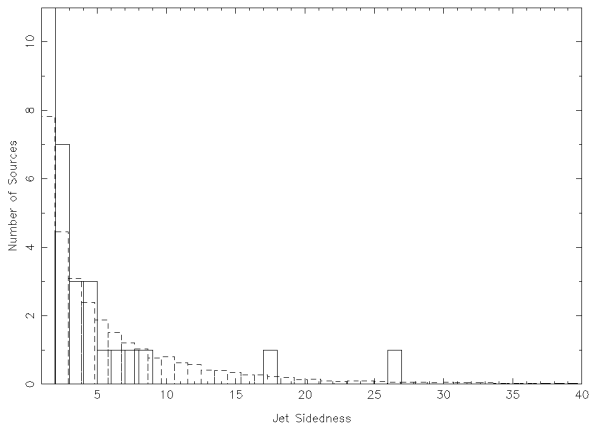}}}
\caption{The simulated data (dashed lines) over-plotted on the sample
  data (solid lines).  The simulated sample shown in Fig.~(a) includes
  no intrinsic sidedness and uses a best fit jet speed of
  (0.60$\pm$0.02)$c$.  Fig.~(b) shows the results of including some
  intrinsic sidedness in the distribution, with a best fitting
  $\beta=0.5\pm0.2$, as described in the text.}
\end{figure}

Using this method, a mean jet speed and dispersion of
$\beta=0.60\pm0.02$ was obtained for simulated data samples obtained
as described above, assuming that beaming is the sole determinant of
the observed sidedness distribution, and that the jet speed found by
the MCMC algorithm holds exactly for all sources.  Thus, any imperfections in the
real data, such as intrinsic scatter, are not taken into account.  We
discuss the addition of intrinsic sidedness in
Section~\ref{sidednessinc}.  The simulated sample and real data are
shown overplotted in Fig.~\ref{3a}.

\subsection{Including the intrinsic sidedness}
\label{sidednessinc}
We modelled the jets as described in Section~\ref{beamingonly}, but
allowed values of $k$ to be drawn from the ratio of two truncated
Gaussian distributions of full width half maximum (FWHM) derived from
the data.  Fitting the data as above, we obtained a jet speed of
$\beta=0.5\pm0.2$ with the best-fitting $\sigma$ of the Gaussians from
which $k$ was drawn equal to 0.4$\pm$0.2.  The errors quoted
  here are derived from simulations as in Section~\ref{beamingonly}.
The data and simulated sample are shown in Fig.~\ref{3b}.
  
\section{Discussion}
\label{discussion}
\subsection{Are WAT jets beamed?}
\label{beamedjets}
The jet speeds with no intrinsic scatter for WAT sources obtained by
\citet{HS2003} and \citet{ODonoghue} are lower than found here (but
within our errors), but the sample used here is larger, and we have
not preferentially selected sources with jet sidedness close to unity;
inadvertent selection of sources with sidedness values close to unity
has hampered previous studies.  Comparing our results to jet speeds
obtained for both FRI and FRII type sources, it appears that the jet
speed results presented here for WAT sources are consistent with mean
jet speeds found for FRI radio sources, at the points where FRI jets
flare ($\left(0.54\pm0.03\right)c$ \citep[][and references
therein]{2004MNRAS.351..727A}).  Using different samples and analysis
methods, \citet{2004MNRAS.351..727A}, \citet{Hardcastle1999} and
\citet{1997MNRAS.286..425W} find speeds in FRII jets in the range
$0.5-0.7c$, which are also consistent with the jet speeds found in
WATs.

There is however, a fundamental degeneracy between the beaming speed
and intrinsic scatter, since we cannot, a priori, rule out a model
that contains no beaming, but has just the right amount of intrinsic
sidedness to reproduce the observed distribution.  However, since
models with and without intrinsic scatter require some beaming, it is
likely that jets in WATs are at least mildly relativistic, with the
beaming speed obtained in this study being an upper limit, as
intrinsic sidedness reduces the required beaming speed.

\begin{figure}
\scalebox{.45}{\includegraphics{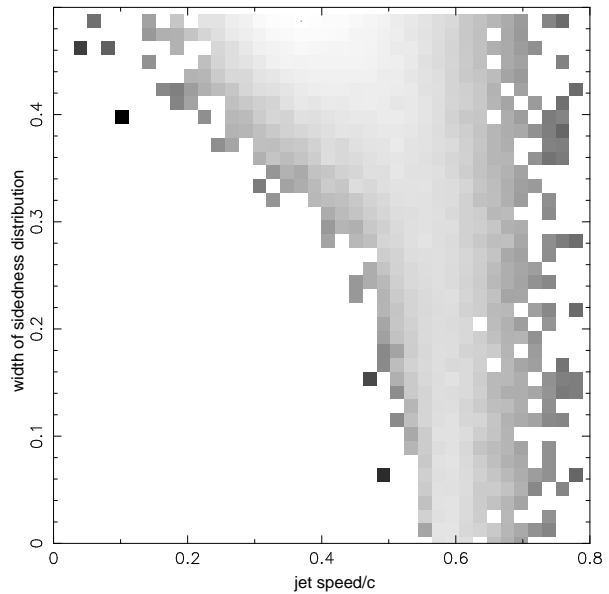}}
\caption{A plot showing the relative likelihood of the different pairs
  of jet speed and jet sidedness values that were fitted to the data.
  The grey-scale indicates the likelihood of each combination, with
  the darkest grey representing those values that are the least
  likely, and the lightest shades of grey representing the most
  likely.  The unshaded areas show regions of the parameter space that
  were not sampled, as they were not expected to yield sensible fits.
  There is very little difference between the likelihood of a model
  with some intrinsic scatter and some beaming, and one with less
  beaming where intrinsic sidedness determines the distribution.  For
  clarity, whilst the actual parameter space investigated is larger
  than that shown, we restrict this plot to the region of parameter
  space with $\sigma < 0.5$ and $\beta<0.8$.}
\label{grid}
\end{figure}

\begin{figure}
\scalebox{.35}{\includegraphics{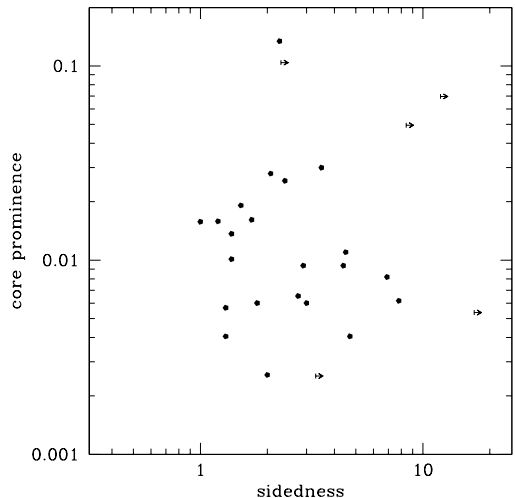}}
\caption{Core prominence plotted against jet sidedness for our WATs
  sample.  The arrows show those points for which limits for
  jet-sidedness were obtained.}
\label{coreprom}
\end{figure}

\citet{1982MNRAS.200.1067O} show that as the core of a radio galaxy is
essentially the unresolved bases of the two relativistic jets, the
radio core should show some evidence for beaming; the core
prominence distribution for a sample of sources, i.e. the distribution
of the ratios of the flux of the beamed core to the flux from the
unbeamed components, should be dependent on the jet speed.  Therefore,
as the jet sidedness distribution is also a function of jet speed, jet
sidedness and core prominence should show some correlation.  However,
a plot of these two quantities (Fig.~\ref{coreprom}) shows at best a
weak correlation.  It would appear that despite the jet speeds found
in Sections~\ref{beamingonly} and \ref{sidednessinc}, the cores in
these sources show only marginal evidence for beaming.  This could
simply indicate that core prominence is not a good indicator of
beaming in these sources, as the jet sidedness ratios do indicate that
at least some relativistic beaming is present.

The use of core prominence as a beaming indicator relies on the
hypothesis that the beamed and unbeamed components of a radio source
are always identical; the only difference between different radio
sources is the angle that the jets make with the line of sight. If
this is not the case, then there will be scatter in the relationship
between jet-sidedness and core prominence, as they will not be
measuring the same thing.  The extended emission in WATs should be
dominated by old material flowing down the lobes.  Therefore, we
expect that the ratio of the unbeamed to the beamed flux of a given
source will depend on the age of the source, weakening the
relationship between core prominence and jet sidedness, as an older
source may still be beamed, but due to a large amount of old material
in the lobes, have a small core prominence.  This age dependence of
unbeamed flux, combined with the scatter in jet speed, could lead to
the core prominence being an unreliable indicator of beaming in WAT
sources.

Given that some WAT sources exhibit hotspot like features
\citep{HS2003}, it would be surprising if at least some relativistic
beaming were not responsible for the observed sidedness distribution.
If the hotspots in WATS are analogous to FRII hotspots, this implies
that WAT hotspots are jet termination shocks.  If that is the case,
then the hotspots indicate that the jet is faster than the internal
sound speed ($c/\sqrt 3$ for a relativistic plasma).  Since hotspots
are not a universal feature of WATs, it would appear that the jet
speed in WATs could be approximately the internal sound speed of the
jets (compare $0.60\pm0.02c$ above for jets with no intrinsic
variance, with $0.58c$ as the internal sound speed in a relativistic
plasma), with intrinsic variation in the jet speed accounting for the
presence or absence of hotspots.  If this is the case, then it
  would be expected that approximately half of the sources in the
  sample would show hotspots, and half would not.  As high resolution
  data is required in order to ascertain the presence of a hotspot,
  the \citet{HS2003} sample, which is a sub-sample of the one
  presented in this paper, was examined to obtain the proportion of
  plumes with and without hotspots.  A plume was categorised as having
  a hotspot if it contained a bright, compact structure, coincident
  with the termination point of a jet.  If there was bright, compact
  structure in the plume, but the end of the jet could not be traced
  into the plume, the region was classified as a possible hotspot.  In
  that sample we find that 5 plumes contain definite hotspots, 5
  plumes do not contain hotspots, and 4 plumes contain bright,
  reasonably compact structures in the plumes, that could be
  associated with jet termination (possible hotspots).  This is
  consistent with the suggestion that intrinsic variation in speed could account
  for the presence or absence of hotspots in WAT sources.

\subsection{Jet bending in WATs}
\label{jetbending}
As seen in \citet{HS2003}, most WAT jets show some slight bending of
the jets.  It can be argued that very light, very relativistic jets
should be `stiff' and resistant to bending, requiring unreasonably
high host galaxy speeds.  However, if the jets are moderately
relativistic, as has been argued in Section~\ref{beamedjets}, then
more reasonable galaxy speeds should be required, provided that the
jets are light.  Following the discussion of jet bending in
\citet{H3C465}, an estimate of the maximum velocity of the host galaxy can
be made.  We use Euler's equation in the form
\begin{equation}
\frac{\rho_{j}v^{2}_{j}}{R}=\frac{\rho_{\mathrm{ext}}v^{2}_{g}}{h_j}
\label{Euler}
\end{equation}
\citep[e.g.][]{1985ApJ...295...80O} where $\rho_j$ is the density of the jet, $v_j$ is the speed of the
jet, $R$ is the radius of curvature of the jet, $\rho_{\mathrm{ext}}$
is the density of the inter-galactic medium, and $h_j$ is the width of
the jet.  We assume a jet speed of $0.6c$, and as in \citet{H3C465} we
take the minimum equivalent external density to be given by $\rho_j =
3p_{\mathrm{min}}/c^2$ where $p_{\mathrm{min}}$ is the minimum
internal pressure of the jet.  Using a proton number density ($n_p$)
of 1500~m$^{-3}$, and the minimum internal pressure of
the jet to be $3\times10^{-12}$~Pa \citep[from][]{H3C465}, a
galaxy speed of 140~km s$^{-1}$ is needed to produce a bend in the jet
of 3C465 of one jet width, which is the most that the jets in 3C465
appear to bend over their lengths.  This galaxy speed is plausible for a large central
galaxy sitting in the centre of X-ray emission from a cluster, as
3C465 does, and given that the estimate of $\rho_j$ is a lower limit,
it seems likely that small scale jet bending can occur in WATs in
clusters, if the jet is light.

In addition to WAT sources that have mainly straight jets, with only
small bends, there is also a class of radio galaxies that whilst
exhibiting classical WAT morphology, also have extremely bent inner
jets; the best example of this class of source is 0647+693
\citep{HS2003}, but other sources with a similar morphology are shown
in Appendix~\ref{maps} (e.g. 2151+085).  If the jet speeds are as fast
as we have deduced here, and we assume that, as in 3C465 above, the
jet bending is caused by the bulk motions of the host galaxy through
the IGM, we can ask whether the galaxy velocity is plausible for a
galaxy sitting at or near the centre of a cluster.

To induce bending of the scale seen in 0647+693, where the distance
from the core to the base of the plume is 30~arcsec \citep{HS2003},
and the jet bends by 20~arcsec over the length of the jet, higher
speeds are required.  Assuming that the jets in 0647+693 have a
similar density to those in 3C465, and that the cluster environment of
0647+693 is similar to that of 2236-176 \citep{Jethaa} as the
temperatures are reasonably similar, the electron number density
($n_e$) at the base of the plume in 0647+693 (30~\kpc from the radio
core) is approximately 1500~m$^{-3}$.  Further assuming that the
proton number density, $n_e=1.18n_p$, an external density at the base
of the plumes of $2.13\times 10^{-24}$~kg~m$^{-3}$ is obtained.
Assuming that $h_j\sim$~1~arcsec, the jet bends by approximately 20
jet widths before the jet enters the base of the plume. Using
Eqn.~\ref{Euler}, a galaxy speed of 870~km s$^{-1}$ is needed to bend
the jets on the scales seen.  This velocity is rather high for a
galaxy near the centre of a cluster, but, if the cluster had undergone
a recent merger event, this velocity would certainly be plausible,
suggesting that WATs with bent jets could be tracing a recent cluster
merger event.  The velocities required to bend the jets also provide
evidence for a model in which the jets are relativistic but light;
jets with densities much higher than the minimum inferred densities
would require much larger velocities to bend.

The same host galaxy speed must also bend the plumes
\citep[e.g.][]{EBOO84}.  Since the plumes, on average, tend to be more
bent than the jets in WATs, this implies that the flow in the plumes
is significantly different to the flow in the jets.  It is already
established that whilst WAT jets may be one-sided, WAT plumes are
always two-sided \citep[3C465 is a good example of this
behaviour][]{H3C465}.  This points to the flow in the plumes being
substantially slower than that in the jets, such that the jets may be
beamed as discussed in this paper, but the plumes certainly will not
be.  This implies that the deceleration of WAT jets may be quite
sudden, and is consistent with observations showing a hotspot-like
feature at the jet termination point.

\section{Conclusions}

We have presented a sample of 30 WATs that we have used to obtain a
representative jet speed for WAT jets.  If we assume that there is no
beaming, then the observed distribution can be reproduced with some
(large) amount of intrinsic sidedness in the jets.  However, if we
assume that the observed sidedness distribution is due to beaming
coupled with some sensible moderate intrinsic distribution, then we obtain a jet
speed in the range $(0.3-0.7)c$.  Whilst the sidedness analysis of the
kpc-scale jets in our WAT sample alone cannot confirm these jet
speeds, comparisons with the jet speeds of FRI and FRII sources
suggests that moderate beaming is at least necessary.  Further, if we
assume that the jets are fast, then our jet bending analysis in
Section~\ref{jetbending}, supports the hypothesis that jets in WATs
are light so that they can be bent through the angles seen in sources
such as 0647+693 whilst requiring reasonable galaxy speeds.

The arguments presented in Section~\ref{beamedjets} outline why core
prominence may not be a reliable indicator of beaming in WAT sources,
and this could explain why the correlation between jet-sidedness and
core prominence is weak in this sample of sources.

Studies of FRI and FRII radio galaxies have established the
relativistic nature of the jets in these sources by examining the
nature of the pc-scale jets, and it is thought that any sidedness that
exists on the pc-scale will persist to the kpc-scale, if the kpc-scale
jets are relativistic.  By examining the pc-scale jets of WATs it
could be established whether the pc-scale jets are relativistic, and
if this persists in the kpc-scale jets.  Some studies of the jets in
WAT sources have been done (e.g. for 3C465 and 0836+290
\citealt{1995ApJ...454..735V} and \citealt{2001ApJ...552..508G}) where
one-sided pc-scale jets are found that are well aligned with the 
kpc-jets.  No proper motions are found as yet, but from these two
objects, it would seem that the one-sidedness persists from pc to
kpc-scale.  This implies that if the pc-scale jet is beamed, then the
kpc-scale jet must also be beamed.  A larger investigation of the
pc-scale jets in for example, the sample of WATs presented here, would
enable us to establish whether the sidedness persists on pc and
kpc-scales for all sources, or just for some.  This would allow the
jet speeds to be further constrained.

\section*{Acknowledgements}
NNJ thanks PPARC for a research studentship.  MJH thanks the Royal
Society for a research fellowship.  IS acknowledges the support of the
European Community under a Marie Curie Intra-European Fellowship.  We
thank Mike Hobson for the use of his {\sc metro} MCMC sampler and the
referee for helpful comments that were useful in improving the paper.
The National Radio Astronomy Observatory is a facility of the National
Science Foundation operated under cooperative agreement by Associated
Universities, Inc.

\bibliographystyle{mn2e} 
\bibliography{MF1419rv2.bib}

\appendix
\section{New radio maps of WATs}
\label{maps}
\begin{figure}
\scalebox{.4}{\includegraphics[angle=270]{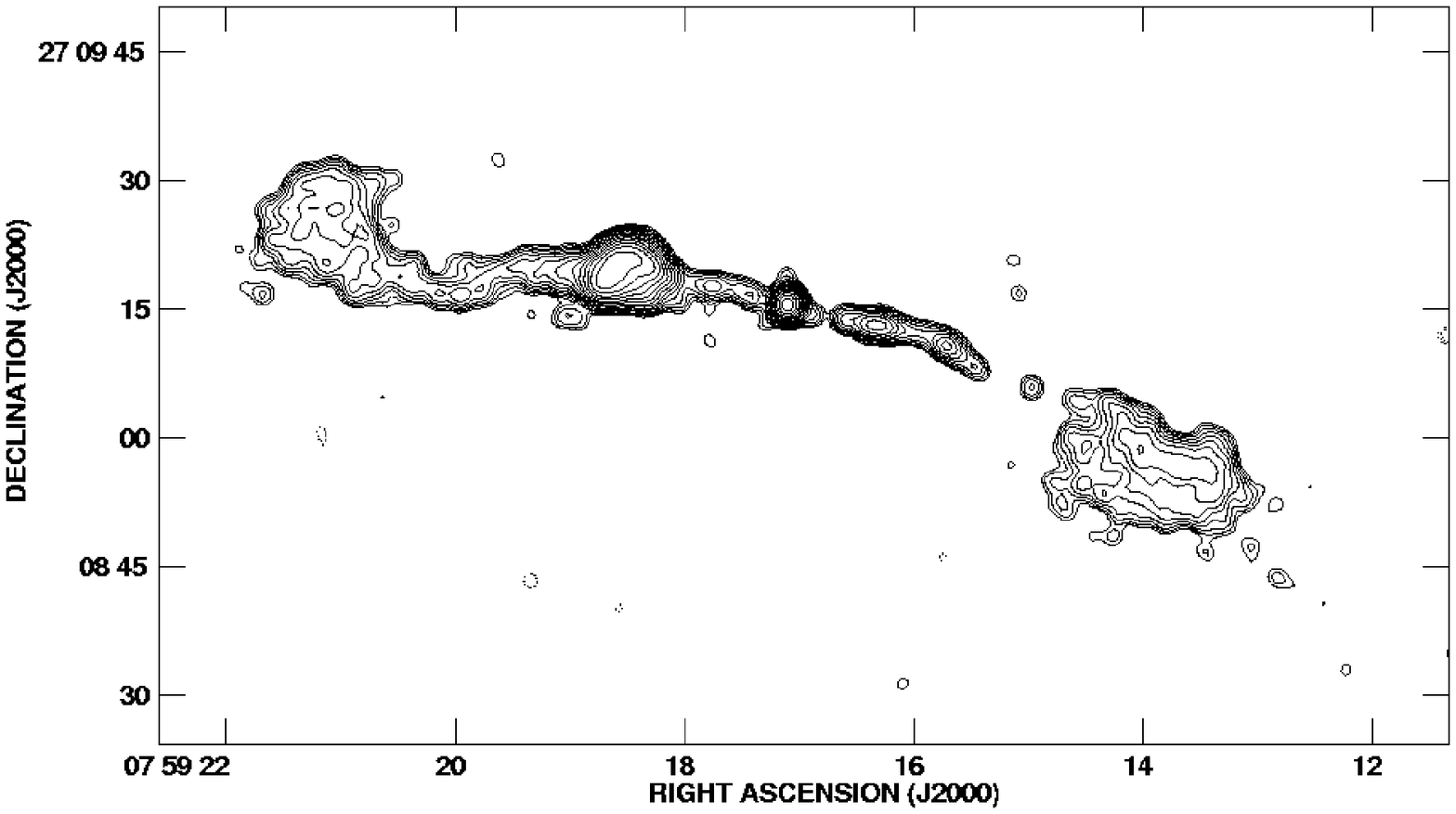}}
\caption{8-GHz, 2.20$\times$2.05-arcsec resolution radio map of 0756+272, hosted by the cluster Abell~610.
  Contours are in logarithmic $\sqrt{2}$ steps, with the lowest
  contour being at 0.25\mJpbeam}
\label{A690}
\end{figure}

\begin{figure}
\scalebox{.4}{\includegraphics[angle=270]{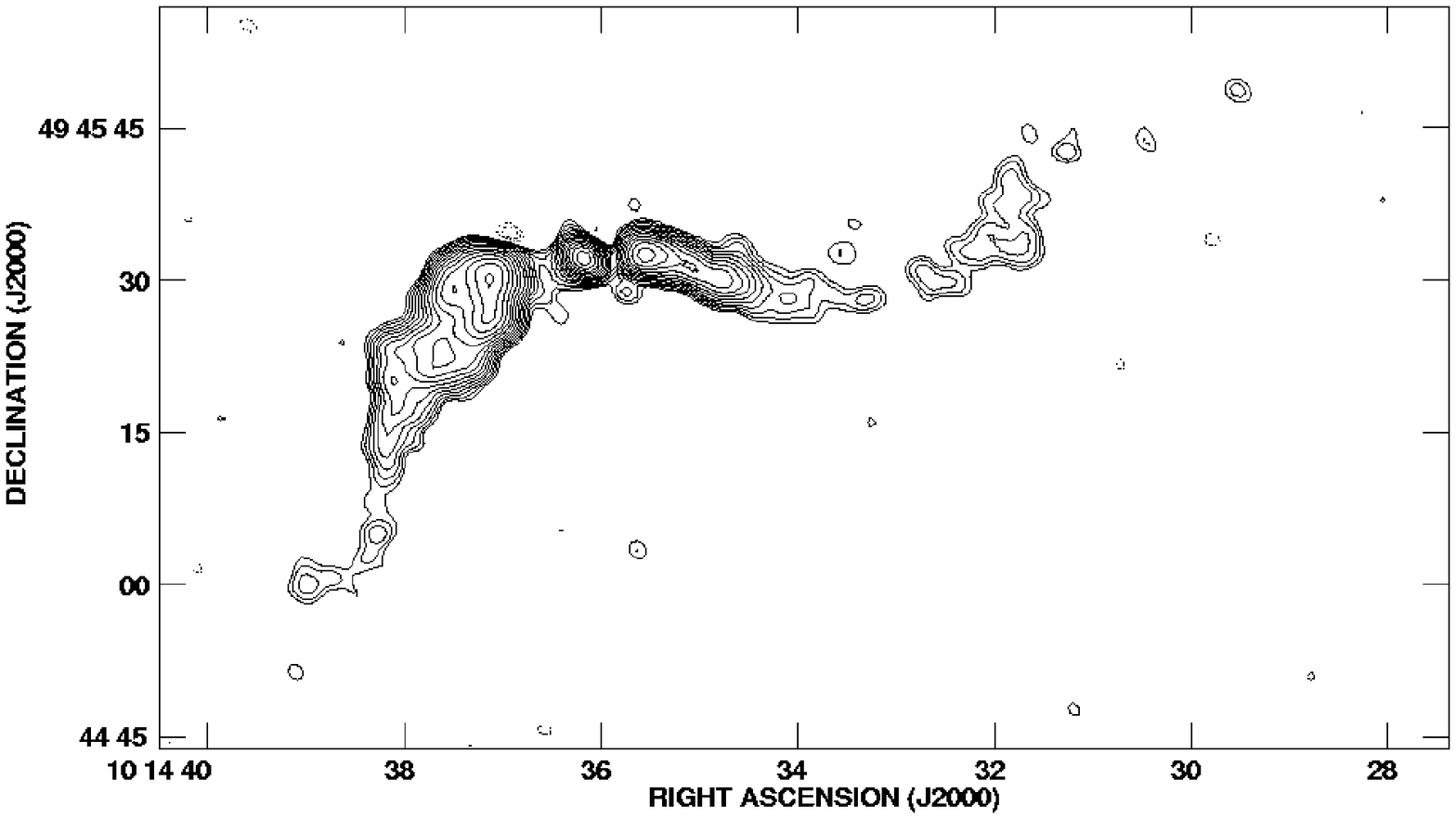}}
\caption{8-GHz, 2.56$\times$2.03-arcsec resolution radio map of 1011+500, hosted by the cluster Abell~950.
  Contours are as Fig~\ref{A690}, with the lowest
  contour being at 0.38\mJpbeam}
\label{A950}
\end{figure}

\begin{figure}
\scalebox{.4}{\includegraphics[angle=270]{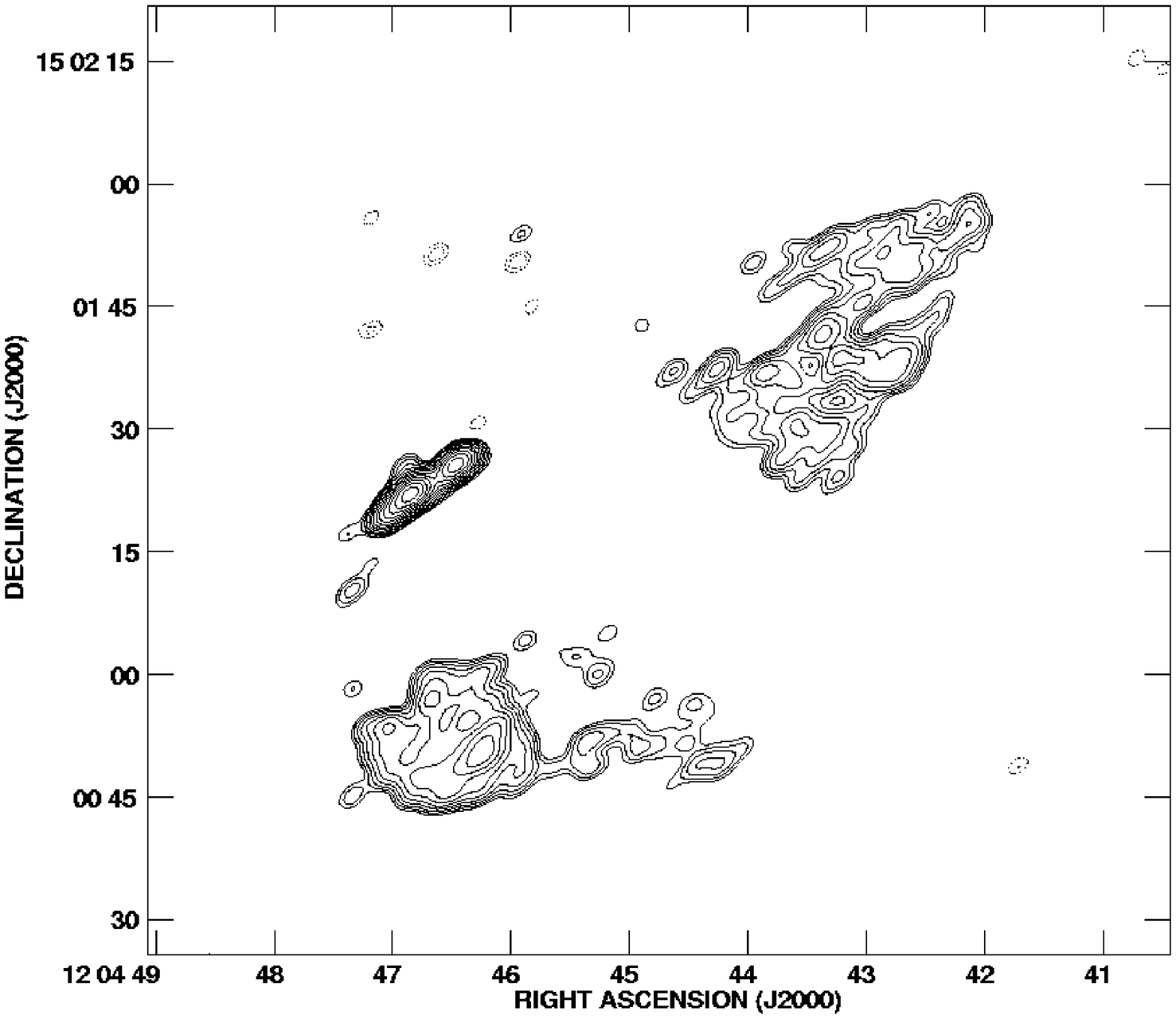}}
\caption{8-GHz, 3.63$\times$2.21-arcsec resolution radio map of 1202+153, hosted by the cluster Abell~1462.
  Contours are  as Fig~\ref{A690}, with the lowest
  contour being at 0.35\mJpbeam}
\label{A1462}
\end{figure}

\begin{figure}
\scalebox{.4}{\includegraphics[angle=270]{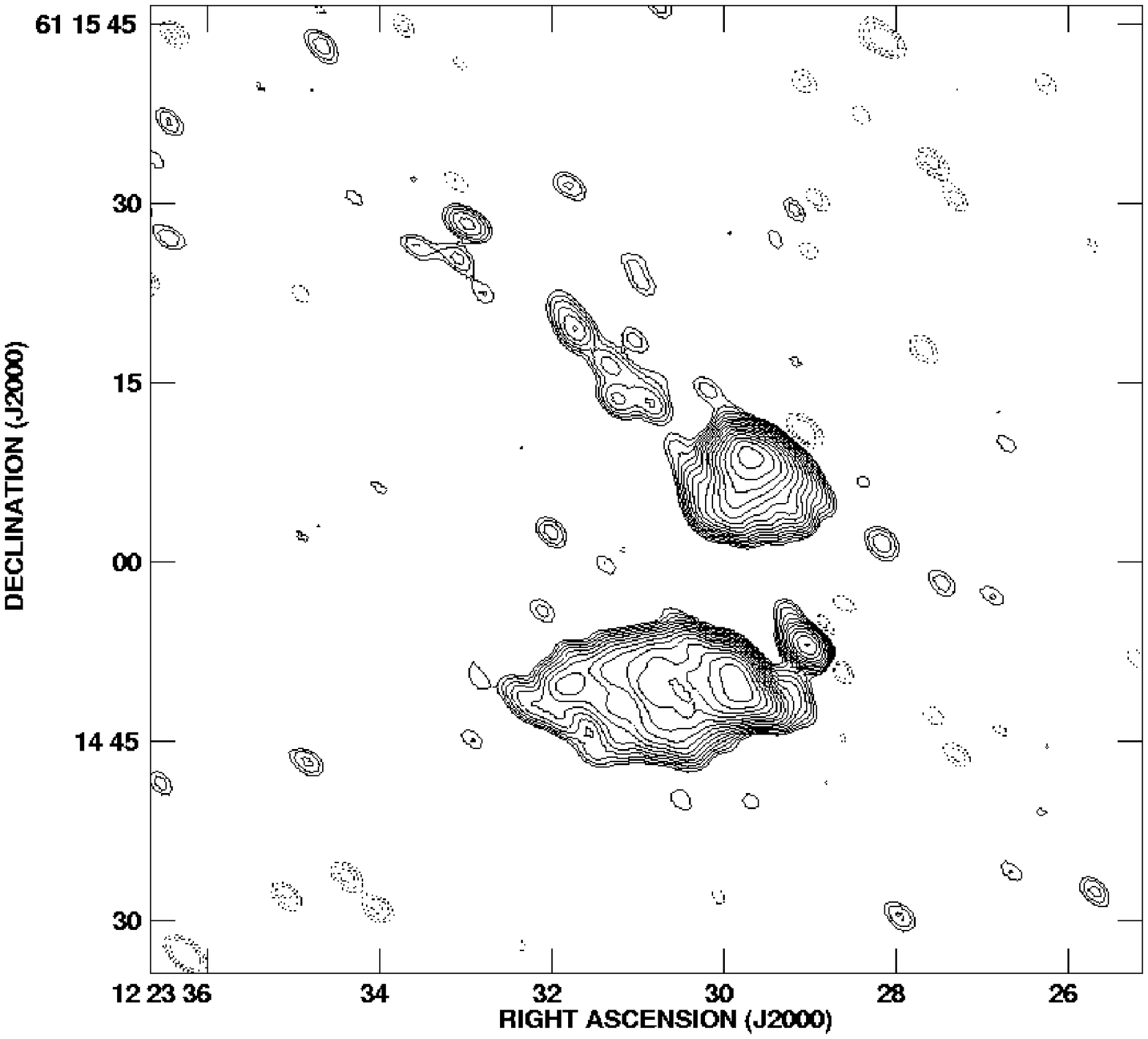}}
\caption{8-GHz, 2.68$\times$1.74-arcsec resolution radio map of 1221+615, hosted by the cluster Abell~1529.
  Contours are  as Fig~\ref{A690}, with the lowest
  contour being at 0.47\mJpbeam}
\label{A1529}
\end{figure}

\begin{figure}
\scalebox{.4}{\includegraphics[angle=270]{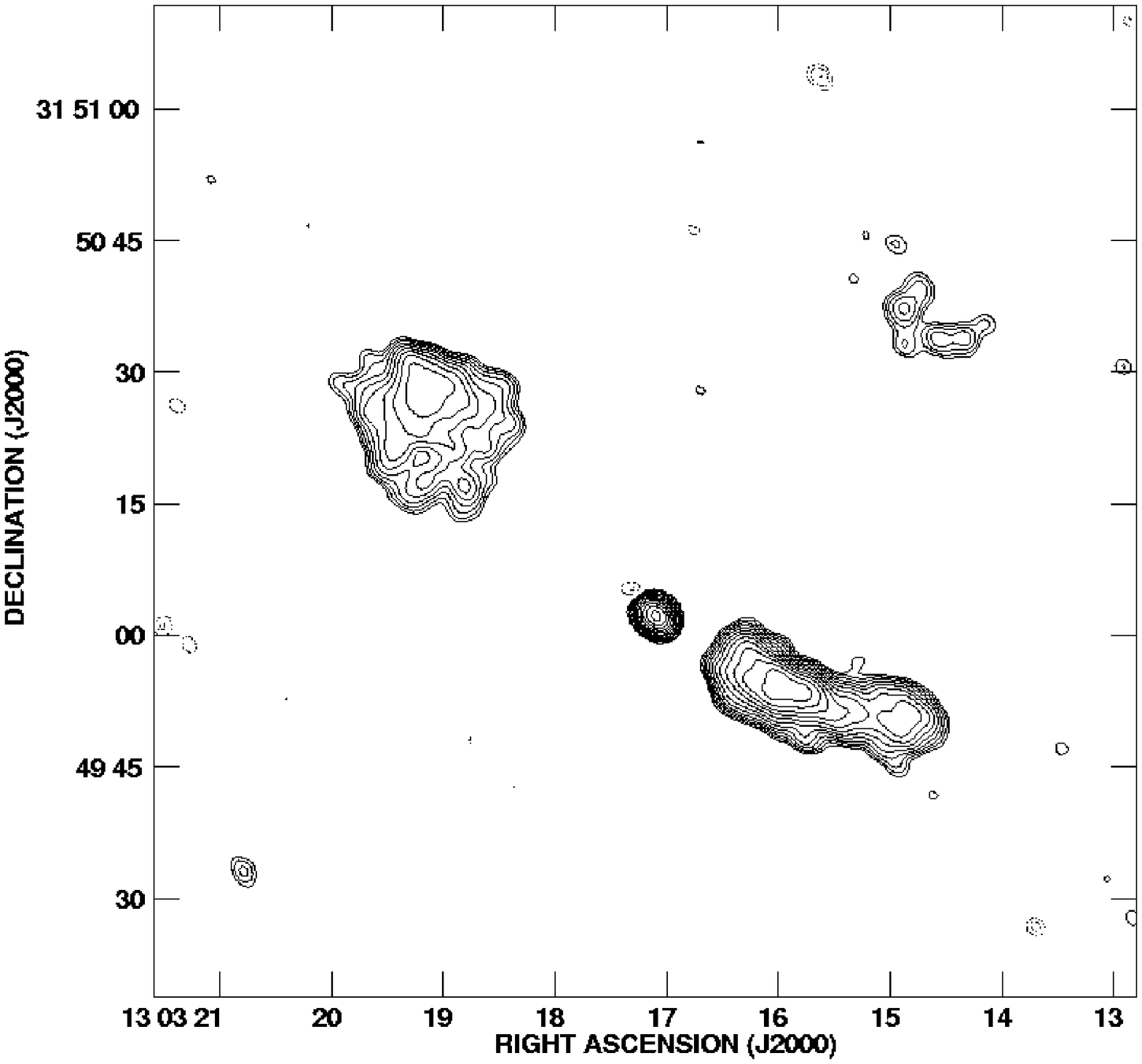}}
\caption{8-GHz, 2.98$\times$2.27-arcsec resolution radio map of 1300+321, hosted by the cluster Abell~1667.
  Contours are  as Fig~\ref{A690}, with the lowest
  contour being at 1.4\mJpbeam}
\label{A1667}
\end{figure}

\begin{figure}
\scalebox{.4}{\includegraphics[angle=270]{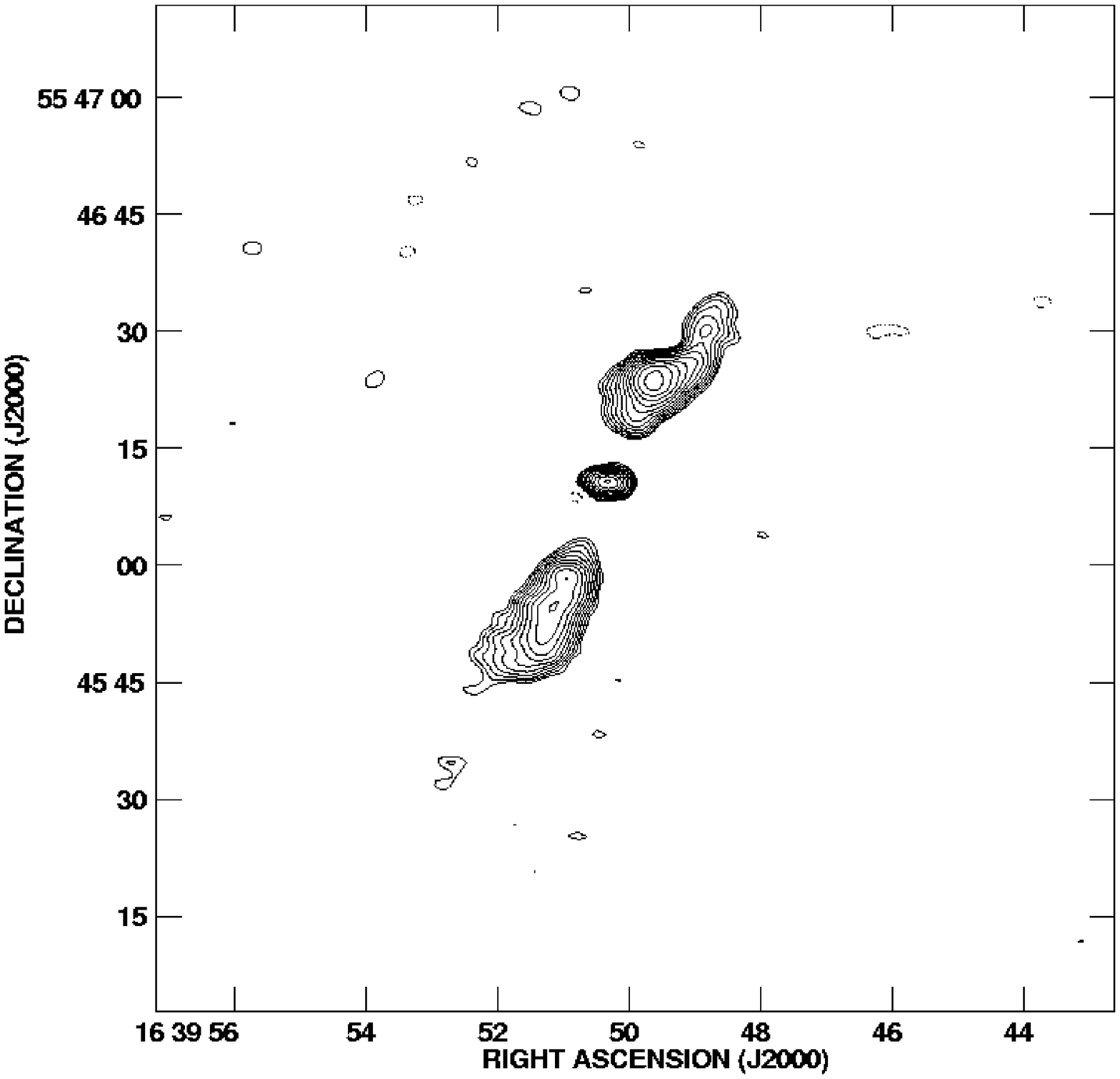}}
\caption{8-GHz, 3.39$\times$2.02-arcsec resolution radio map of 1638+558, hosted by the cluster Abell~2225.
  Contours are  as Fig~\ref{A690}, with the lowest
  contour being at 0.012\mJpbeam.  Whilst we cannot detect the jets in
this source, it is still part of the WAT sample, but was not used in
the jet-speed analysis.}
\label{A2225}
\end{figure}

\begin{figure}
\scalebox{.4}{\includegraphics[angle=270]{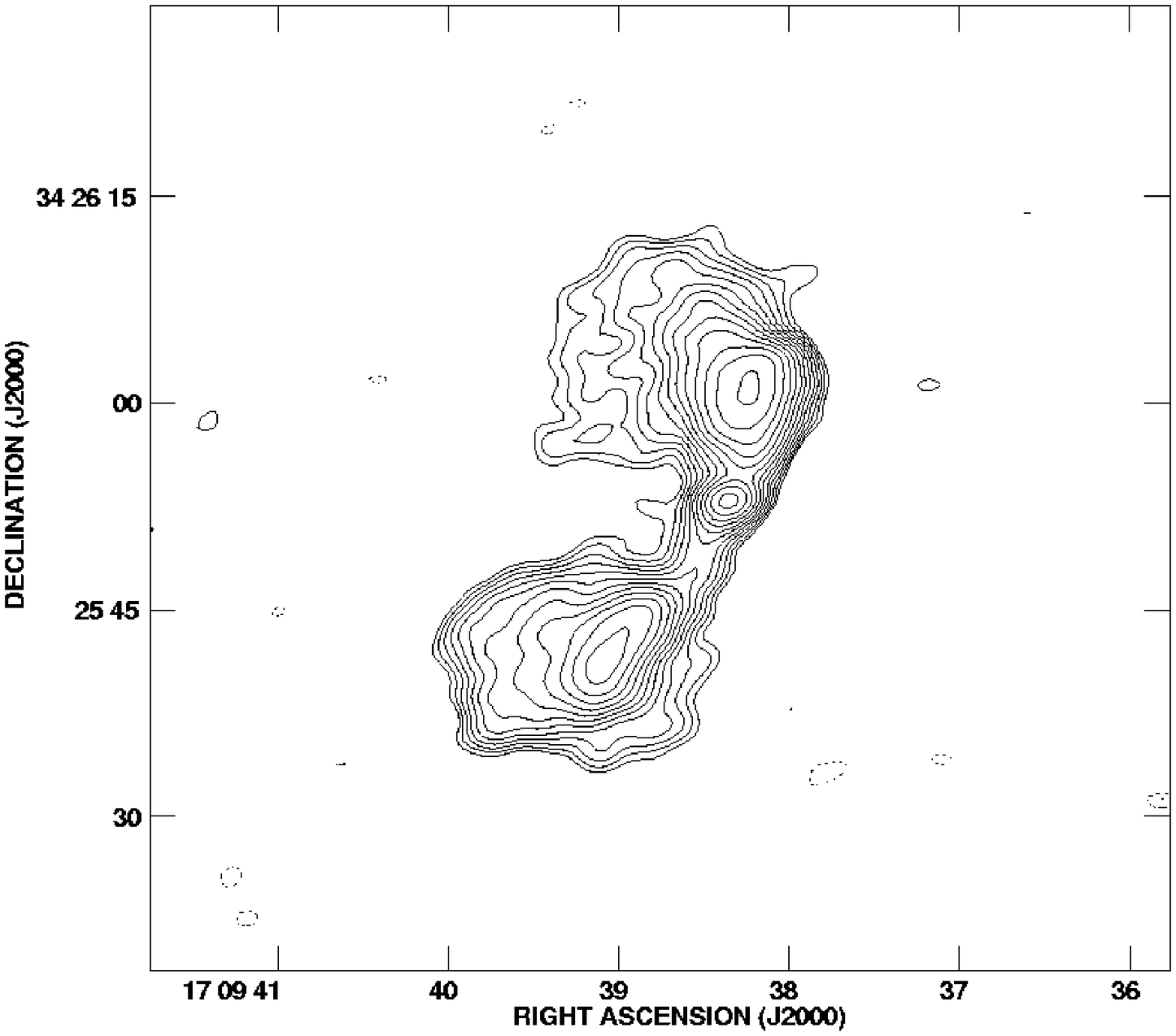}}
\caption{8-GHz, 2.91$\times$1.94-arcsec resolution radio map of 1707+344, hosted by the cluster Abell~2249.
  Contours are as Fig~\ref{A690}, with the lowest
  contour being at 0.20\mJpbeam}
\label{A2249}
\end{figure}

\begin{figure}
\scalebox{.4}{\includegraphics[angle=270]{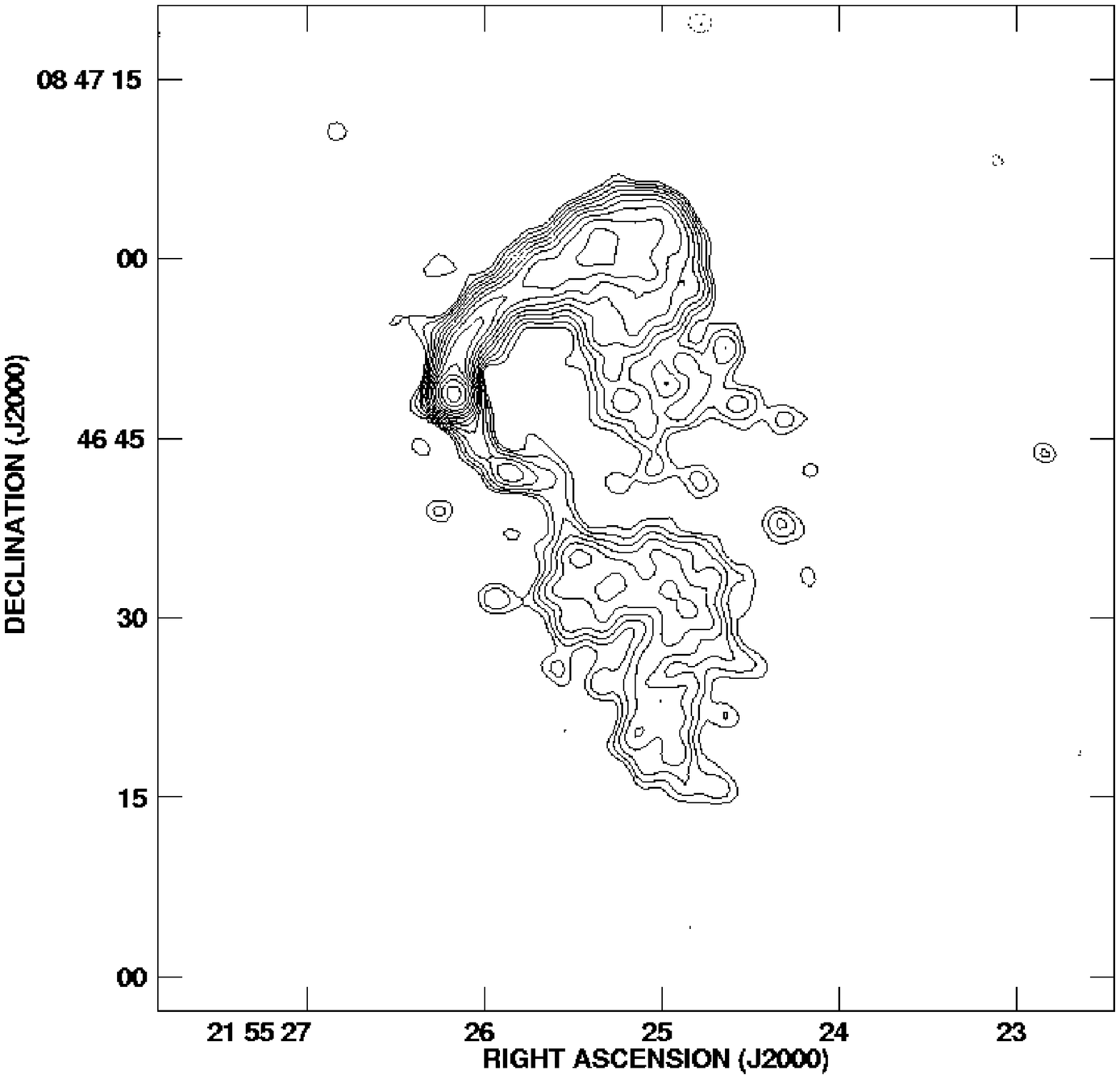}}
\caption{8-GHz, 2.22$\times$2.06-arcsec resolution radio map of 2151+085, hosted by the cluster Abell~2395.
  Contours are  as Fig~\ref{A690}, with the lowest
  contour being at 0.080\mJpbeam}
\label{A2395}
\end{figure}

\begin{figure}
\scalebox{.4}{\includegraphics[angle=270]{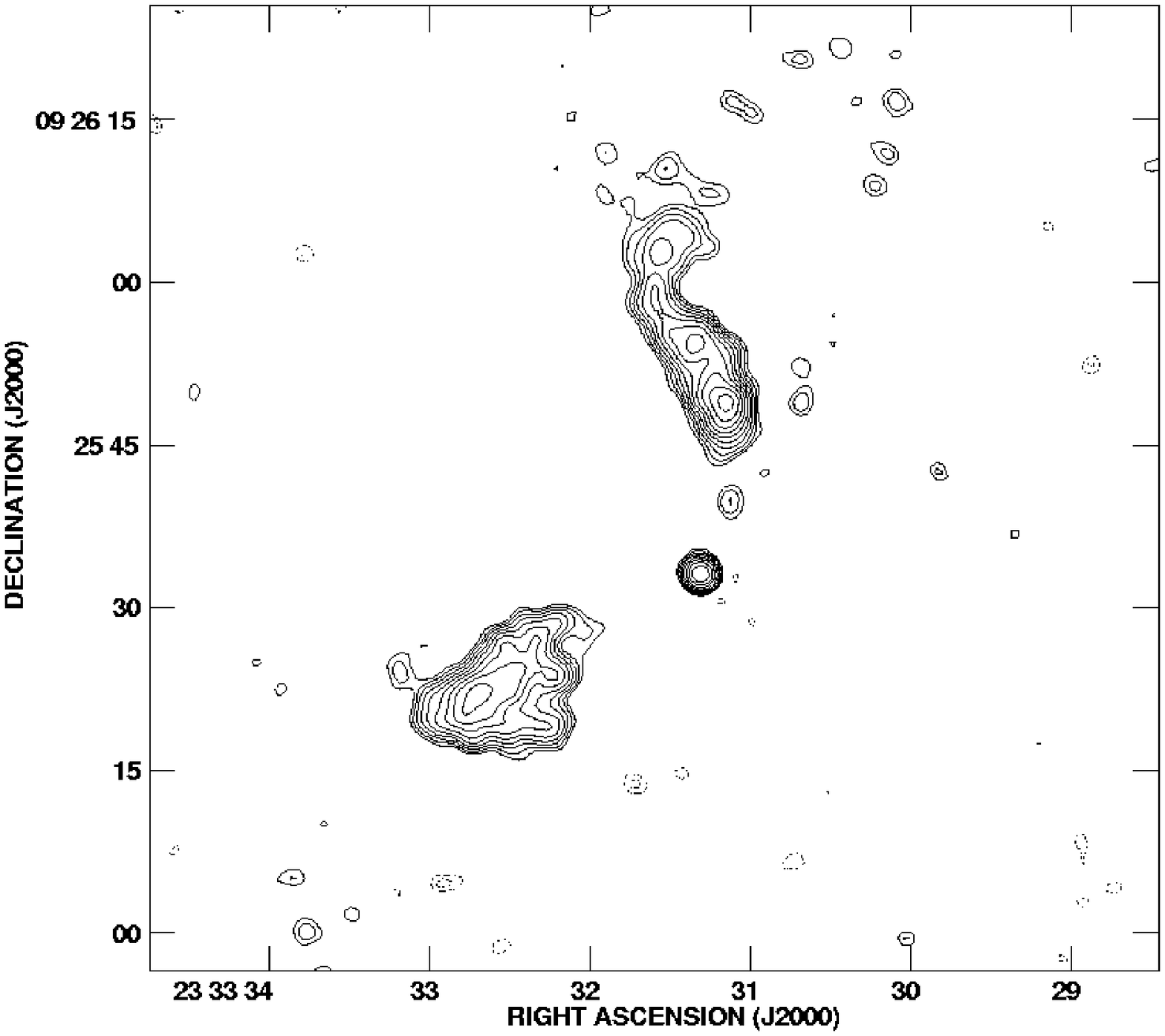}}
\caption{8-GHz,2.07$\times$2.02-arcsec resolution radio map of 2330+091, hosted by the cluster Abell~2617.
  Contours are  as Fig~\ref{A690}, with the lowest
  contour being at 0.073\mJpbeam.  Whilst we cannot detect the jets in
this source, it is still part of the WAT sample, but was not used in
the jet-speed analysis.}
\label{A2617}
\end{figure}

As discussed in Section~\ref{sample}, new data for some of our sample
were obtained, in order to enlarge the WATs sample, where no good
quality data were available.  Here we present the new radio maps
obtained during our VLA observing programme (Figs~\ref{A690} -
\ref{A2617}).  We also present some previously unpublished maps from
data taken from the VLA archive.  In Table~\ref{newobs} we show the
details of the observations used to make the maps presented here.

\begin{table*}
\caption{Details of observations for which new maps are presented in Appendix~\ref{maps}}
\begin{tabular}{lcccclll}
Abell cluster & radio name & array         & observation & time on & observation & map resolution & VLA proposal\\ 
              & (B1950)    & configuration & frequency   & source  & date        &                & number\\ 
              &            &               & (GHz)       &   (min) &             & (arcsec$\times$arcsec)&\\\hline
A610          &0756+272    &  C            & 8.5         &39       &2004-MAR-30  &2.20$\times$2.05& AJ309\\
A950          &1011+500    &  C            & 8.5         &39       &2004-MAR-30  &2.56$\times$2.03& AJ309\\
A1462         &1202+153    &  C            & 8.5         &39       &2004-MAR-30  &3.63$\times$2.21& AJ309\\
              &            &  C            & 8.5         &28       &2004-APR-26  && AJ309\\
A1529         &1221+615    &  C            & 8.5         &39       &2004-MAR-30  &2.68$\times$1.74& AJ309\\ 
              &            &  C            & 8.5         &29       &2004-APR-26  &\\
A1667         &1300+321    &  C            & 8.5         &39       &2004-MAR-30  &2.98$\times$2.27& AJ309\\
A2225         &1638+558    &  BC           & 8.5         &48       &2004-FEB-2   &3.39$\times$2.02& AJ309\\
              &            &  C            & 8.5         &31       &2004-APR-26  && AJ309\\
A2249         &1707+344    &  BC           & 8.5         &45       &2004-FEB-2   &2.91$\times$1.94& AJ309\\
              &            &  C            & 8.5         &31       &2004-APR-26  && AJ309\\
A2395         &2151+085    &  BC           & 8.5         &47       &2004-FEB-2   &2.22$\times$2.06& AJ309\\
A2617         &2330+091    &  BC           & 8.5         &48       &2004-FEB-2   &2.07$\times$2.02& AJ309\\
A1446         &1159+583    &  B            & 4.86        &125      &1984-FEB-17  &1.29$\times$ 1.06 & AO049\\
A2029         &1508+059    &  A            & 8.5         &367      &1992-NOV-14  &0.29$\times$0.26 & AT144\\
              &            &  B            & 8.5         &151      &1993-MAR-24  &                 &\\
A1684         &1306+107A   &  B            & 4.86        &136      &1985-DEC-07  &13.25$\times$11.88  & AO062\\
A194          &3C40        &  B            & 1.5         &335      &1984-JAN-08  &5.58$\times$5.17 & AV102\\
              &            &  C            & 1.5         &206      &1984-JUN-02  &                 & AV102 \\
              &            &  D            & 1.5         &81       &1984-JUL-31  &                 & AV112  \\
\end{tabular}
\label{newobs}
\end{table*}

\begin{figure}
\scalebox{.4}{\includegraphics{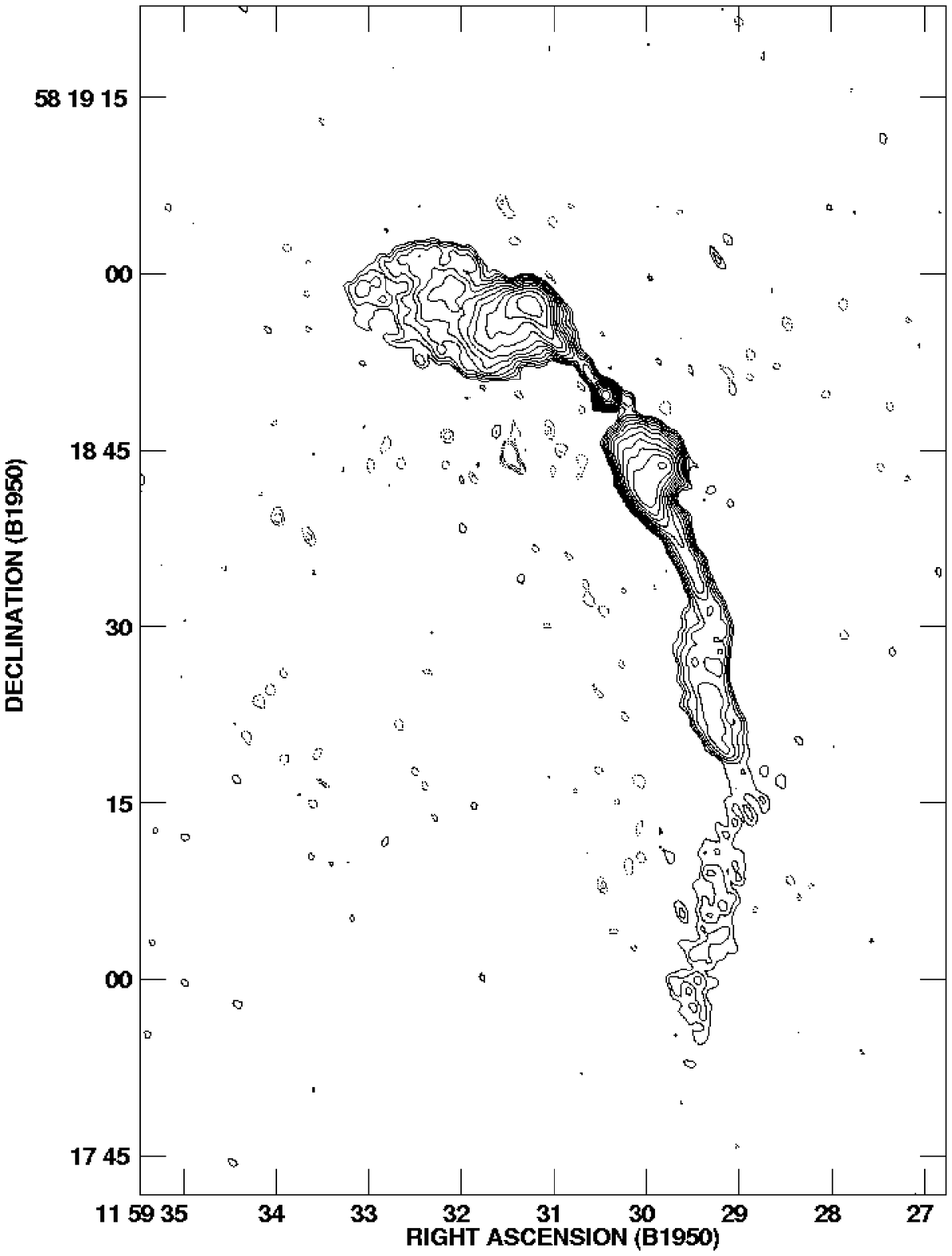}}
\caption{4.86-GHz, 1.29$\times$ 1.06-arcsec resolution radio map of 1159+583, hosted by
  the cluster Abell~1446.  Contours are as Fig~\ref{A690}, with the
  lowest contour at 0.13\mJpbeam.  The map was made with data obtained
  from the VLA archive}
\label{A1446}
\end{figure}

\begin{figure}
\scalebox{.4}{\includegraphics{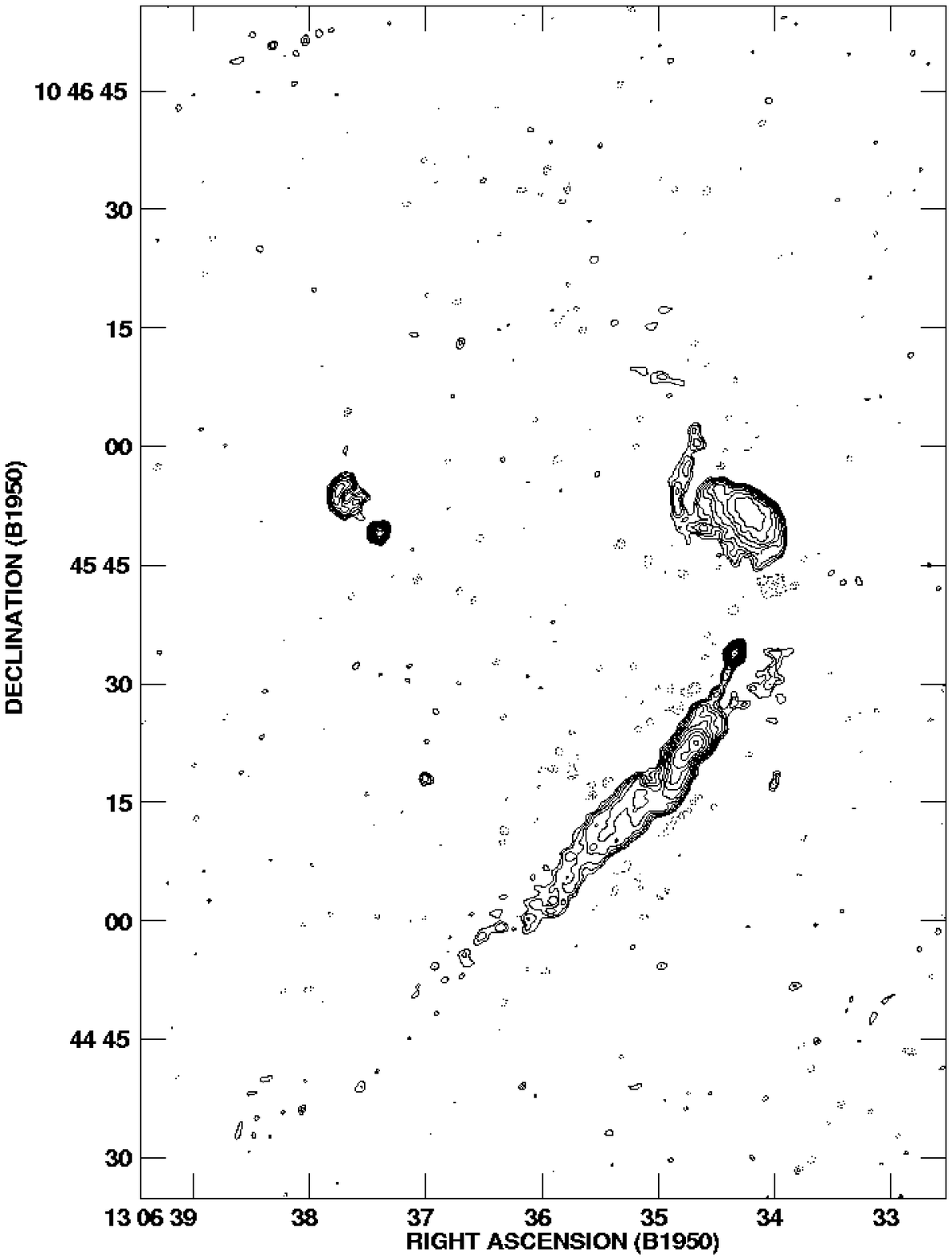}}
\caption{4.86-GHz, 1.29$\times$ 1.06-arcsec resolution radio map of 1306+107A, hosted by
  the cluster Abell~1684.  Contours are as Fig~\ref{A690}, with the
  lowest contour at 0.10\mJpbeam.  The map was made with data obtained
  from the VLA archive}
\label{A1684}
\end{figure}

\begin{figure}
\scalebox{.4}{\includegraphics{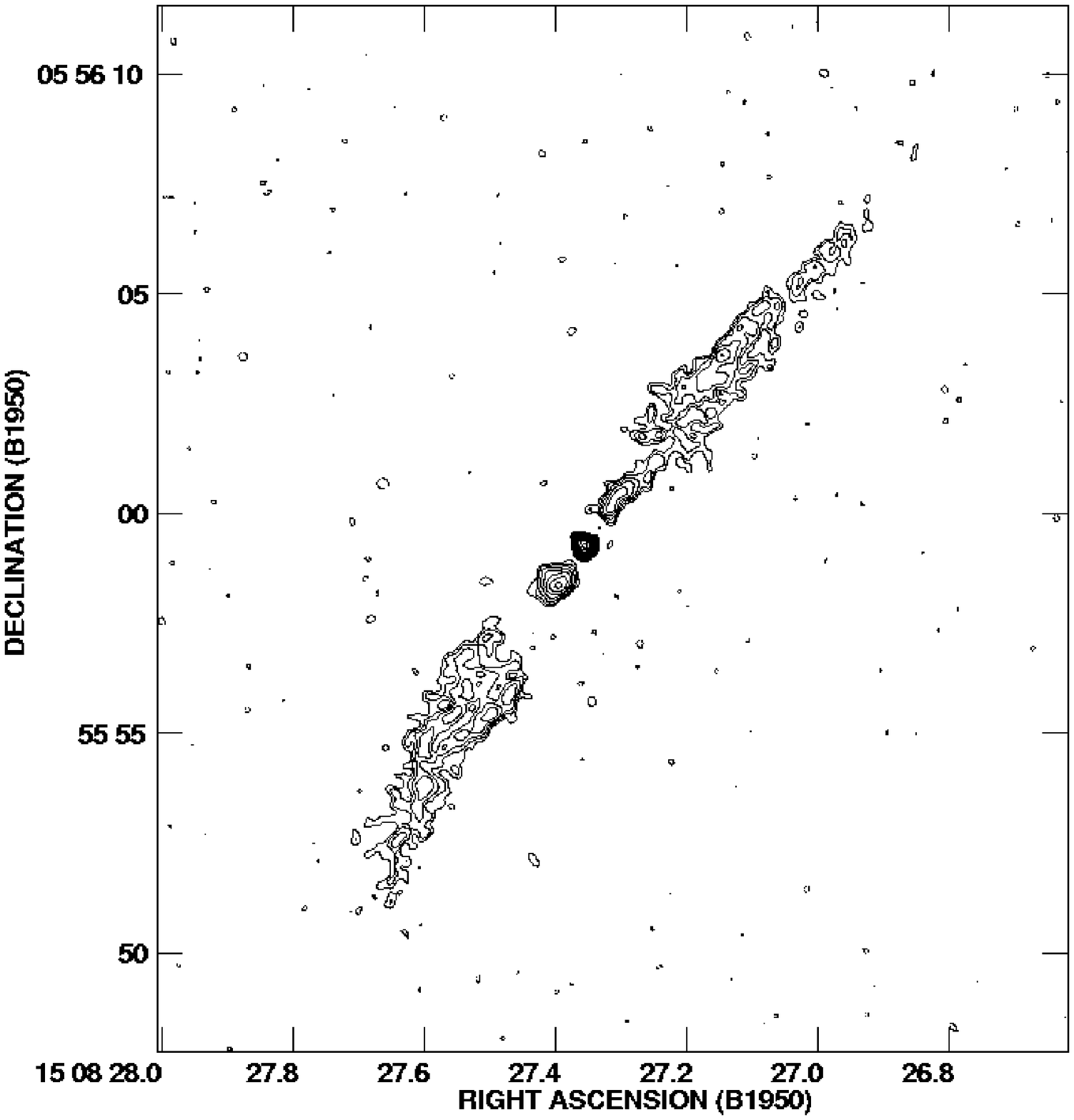}}
\caption{8.0-GHz, 0.29$\times$0.26-arcsec resolution radio map of 1508+059, hosted by
  the cluster Abell~2029.  Contours are as Fig~\ref{A690}, with the
  lowest contour at 0.054\mJpbeam.  The map was made with data obtained
  from the VLA archive}
\label{A2029}
\end{figure}

\begin{figure}
\scalebox{.4}{\includegraphics{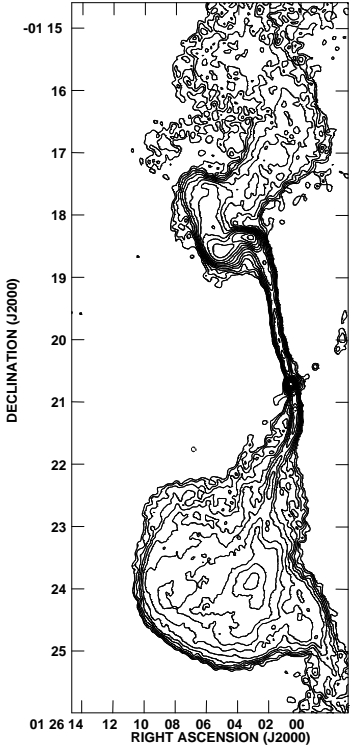}}
\caption{1.5-GHz, 5.58$\times$5.17-arcsec resolution radio map of 3C40, hosted by
  the cluster Abell~194.  Contours are as Fig~\ref{A690}, with the
  lowest contour at 0.40\mJpbeam.  The map was made with data obtained
  from the VLA archive}
\label{A194}
\end{figure}

\end{document}